\documentstyle[prl,aps,epsfig,multicol]{revtex}
\begin{document}
\title{Geometrically-controlled twist transitions in nematic cells}

\author{Pedro Patr\'\i cio $^1$, M.M. Telo da Gama $^1$, 
and S. Dietrich $^{2,3}$} 

\address{$^1$ Departamento de F{\'\i}sica da Faculdade de Ci{\^e}ncias and 
Centro de F{\'\i}sica Te\'orica e Computacional\\
Universidade de Lisboa, 
Avenida Professor Gama Pinto 2, P-1649-003 Lisboa Codex, Portugal}

\address{$^2$Max-Planck-Institut f\"ur Metallforschung, 
Heisenbergstr. 1, D-70569 Stuttgart, Germany}

\address{$^3$Institut f\"ur Theoretische und Angewandte Physik, 
Universit\"at Stuttgart, Pfaffenwaldring 57, D-70569 Stuttgart, Germany}

\date{February 2002}

\maketitle

\begin{abstract}

We study geometrically-controlled twist transitions
of a nematic confined between a sinusoidal grating and a flat substrate.
In these cells the transition to the twisted state is driven by surface
effects.
We have identified the mechanisms responsible for the transition
analytically and used exact numerical calculations to study the range
of surface parameters where the twist instability occurs.
Close to these values the cell operates under {\em minimal} external fields
or temperature variations.

\end{abstract}

\vspace{2mm}

\begin{multicols}{2}

Liquid-crystal (LC) displays consist of a LC layer confined between two
surfaces that impose a preferred orientation of 
the average molecular direction {\bf n}({\bf r}). Current nematic
displays rely on voltage-induced reorientation of the director within the
{\em bulk} LC layer. 
Recently, textured surfaces (on scales of the order of 1$\mu$m) were
designed for patterned alignment of LC \cite{Bryan-Brown,Gupta,Lee,Shah}, 
opening possibilities for an improved performance of LC cells.
In view of the very rich behavior of LCs even on homogeneous substrates
\cite{Jerome} and of simple fluids on structured surfaces \cite{Dietrich}
such technological advances beyond a trial and error procedure require
theoretical guidance. 
As a first step in this direction 
we establish the minimal model appropriate 
to the theoretical description of patterned LC cells and develop an
efficient method of solving it. 
This study contributes to the broader perspective of how
well-defined structures on solid surfaces can be imprinted
on adjacent soft matter. The particular softness of LCs leads
to the expectation of very pronounced effects.

The microscopic description of LC surfaces and interfaces is
complex. The degree of nematic order, biaxility, etc.
vary in the interfacial region over {\em molecular} distances while the
director may vary over {\em macroscopic} distances.
In LC displays, the characteristic distance over which the director 
varies is set by the cell dimensions or by the electric
correlation length \cite{deGennes-Prost}. 
Under most experimental conditions 
the length over which the director varies is of the order of $\mu$m and a
macroscopic or elastic theoretical description is adequate
\cite{deGennes-Prost,Chaikin}. 
The macroscopic (second order) elastic free energy of bulk nematics was 
established more than 40 years ago \cite{Frank} but the status of the
surface contributions in the weak anchoring 
regime is still controversial \cite{Yokoyama,FournierPRL}.

In the following we propose and minimize an elastic free energy
for patterned LC displays that includes bulk and surface terms.
We consider the twist cell proposed in a recent experiment, where the
nematic
is confined between a flat and a sinusoidal 
grating surface \cite{Bryan-Brown}.
In the experiment a voltage-controlled
twist (VCT) effect highly sensitive to the {\em surface} properties of
the cell (grating geometry and anchoring strength) has been reported, 
for gratings on the scale of tenths of $\mu$m. 
Berreman \cite{Berreman} and de Gennes \cite{deGennes-Prost}
first considered grating surfaces to
explain azimuthal anchoring by elastic effects only.
Faetti \cite{Faetti}, and more recently 
Fournier and Galatola \cite{FournierPRE}, generalized the effective 
azimuthal anchoring energy by introducing
local anchoring at the grating surface.
Barbero and Durand \cite{Barbero} also considered grating surfaces 
with characteristic lengths comparable to the nematic correlation
length.
They used the Landau-de Gennes free energy \cite{deGennes-Prost}
to describe the induced quasi-melting caused by the rough surface.
Over the last few years there has been a considerable surge
of interest in the influence of grating-like surfaces on the
structural properties of LCs because
new techniques enabled the manufacture of controlled undulated surfaces,
allowing for meaningful comparisons between theory and experiments.
Among these new systems the aforementioned
twist displays exhibit excellent viewing
angle characteristics \cite{Bryan-Brown}, that are important
for technological applications.

\begin{figure}[ht]
\par\columnwidth=20.5pc
\hsize\columnwidth\global\linewidth\columnwidth   
\displaywidth\columnwidth
\centerline{\epsfxsize=240pt\epsfbox{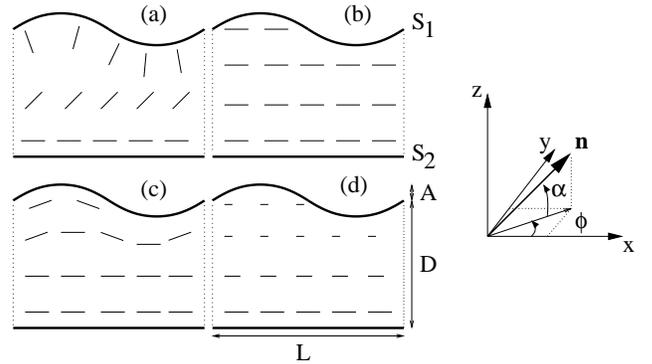}}
\caption
{Nematic director field (projected onto the xz plane) 
for increasing anchoring strengths $W_1$.
(a) Distorted director field, for strongly negative anchoring strength $W_1$. 
(b) Upon increasing $W_1$ a first instability leads to a 
nearly uniform director field.
(c) For larger $W_1$ the director field 
follows the orientation of the surface.
(d) A second instability induces a twisted director field.}
\label{fig:f1}
\end{figure}

In spite of its importance, a theoretical study of twist cells
covering the whole range of operating parameters is still lacking. 
In order to explore how variations in the geometry 
of the cell can induce
the reorientation of the bulk nematic director we consider the total free
energy of the LC cell as the sum
of a bulk elastic free energy $F_b$ and a surface contribution $F_s$. 
The former is the Frank elastic free energy \cite{deGennes-Prost}
\begin{eqnarray}
F_b&=& 
\frac{1}{2} \int_V \{ K_{11} (\nabla \cdot {\bf n})^2 + 
K_{22} [{\bf n} \cdot ( \nabla \times {\bf n)}]^2  \nonumber \\
\label{eq:bulk}
&+&K_{33} [{\bf n} \times ( \nabla \times {\bf n})]^2 \}d^3{\bf r}
\end{eqnarray}
where $K_{11}$, $K_{22}$, and $K_{33}$ are the elastic constants
associated with splay, twist, and bend distortions, respectively.
$F_s$ includes the anchoring energy
for which we adopt the Rapini-Papoular form \cite{Rapini}
\begin{equation}
\label{eq:anchoring}
F_s = 
\frac{W_1}{2} \int_{S_1}  ({\bf n}\cdot{\bf\nu})^2 d^2{\bf r}+
\frac{W_2}{2} \int_{S_2}  ({\bf n}\cdot{\bf\nu})^2 d^2{\bf r} \;.
\end{equation}
These integrals run over the two 
cell surfaces and $W_i$, $i=1,2$, is the corresponding anchoring strength
that characterizes each surface. ${\bf\nu}$ is the local unit vector normal to
the surface. For negative $W_i$ this energy contribution favors normal surface 
orientation of the nematic while positive $W_i$ favor planar (degenerate)
orientation at the surface.
For inhomogeneous substrates in the weak anchoring regime
one may have to include a surface elastic term associated with the
saddle-splay distortion \cite{Yokoyama}. Its contribution will
be considered later.  

\begin{figure}[ht]
\par\columnwidth=20.5pc
\hsize\columnwidth\global\linewidth\columnwidth   
\displaywidth\columnwidth
\centerline{\epsfxsize=200pt\epsfbox{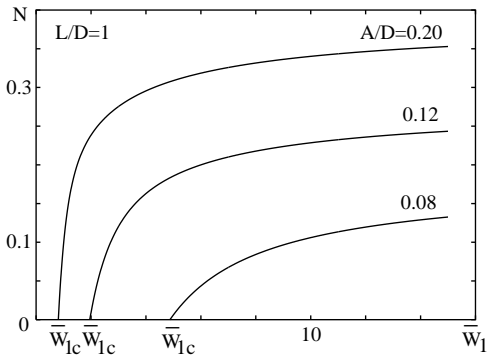}}
\caption
{The twist instability (numerical results).
The twist order parameter $N=D^{-3}\int_V{\bf n}_y^2d^3{\bf r}$
as a function of the reduced anchoring strength 
$\overline W_1=W_1D/K_{11}$ for different groove depth $A/D$.
To favor twist configurations we used $K_{11}=K_{33}=2K_{22}$.
$N$ vanishes linearly at $\overline W_{1c}$.}
\label{fig:f2}
\end{figure}

The total free energy, $F_t=F_b+F_s$, is a functional of the
two angles $\alpha({\bf r})$ and $\phi({\bf r})$, characterizing the nematic
director ${\bf n}=(\cos\alpha\cos\phi,\cos\alpha\sin\phi,\sin\alpha)$.
The cells considered in this letter consist of a nematic LC
confined between a sinusoidal grating 
($z=D+A\sin qx$, where $q=2\pi/L$ and $A$ is the groove depth) and a
flat substrate ($z=0$) (see Fig.~1). 
In Ref.~\cite{Bryan-Brown} the surfaces 
were treated so that the grating induces (weak) homeotropic, i.e., 
normal anchoring with respect to the local surface direction, 
while the flat surface induces (strong) 
homogeneous uniaxial anchoring in the direction perpendicular to
the grooves. In the following calculations we keep the
homogeneous strong anchoring condition at the flat surface 
($\alpha(x,z=0)=\phi(x,z=0)=0$) but we vary
the anchoring strength $W_1$ (including its sign) at the grating surface.
Finally, we use periodic boundary conditions 
$\alpha(x=0,z)=\alpha(x=L,z)$ and $\phi(x=0,z)=\phi(x=L,z)$.

Figure~1 illustrates the behavior of a confined LC with a
small twist elastic constant as the anchoring strength $W_1$ increases.
For large negative values of $W_1$, 
the grating surface induces normal orientation of the nematic.
Upon increasing $W_1$ a
first instability occurs so that the director field becomes
nearly uniform resulting from the competition of two effects: 
the anchoring energy favoring homeotropic anchoring, and the
elastic energy favoring homogeneous alignment 
at the grated surface.
The critical anchoring strength 
for a rectangular cell $D\times S$, 
where $\alpha(x,z)=\alpha z/D$, is found easily.
For small $\alpha$, the total free energy is 
$F_t=(S/2)(W_1+K_{11}/D)\alpha^2$, where $S$ is the area of the 
flat cell surface.
The director field is uniform when the coefficient of $\alpha^2$ is
positive, i.e., for $W_1>-K_{11}/D$.
In the limit of small groove depth $A$ the corrections
arising from the grating are found by first-order perturbation analysis
about the rectangular cell.
When $W_1$ becomes positive, the nematic orientation changes
continuously to follow the sinusoidal shape of the boundary.
Beyond a certain threshold, the bulk bending energy of this deformation
is comparable with a bulk twist deformation  
and a new instability occurs.
Owing to the degeneracy of the planar anchoring,
the sinusoidal and the twisted configuration have approximately the same
surface free energy.

\begin{figure}[ht]
\par\columnwidth=20.5pc
\hsize\columnwidth\global\linewidth\columnwidth   
\displaywidth\columnwidth
\centerline{\epsfxsize=200pt\epsfbox{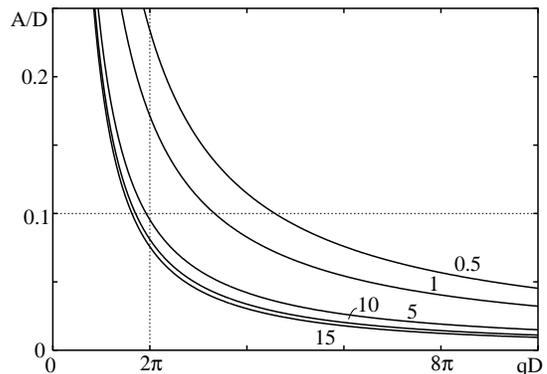}}
\caption
{The onset of the twist instability occurs for anchoring strengths
$W_{1c}$ which depend on the cell geometry $(A,q=2\pi/L)$.
Here the contour lines for $\overline W_{1c}=0.5,1,5,10$, and $15$
are shown for $K_{11}=K_{33}=2K_{22}$.
These results are obtained analytically based
on Eq.(\ref{eq:twist_threshold}).
The vertical and horizontal line indicates systems
with $L=D$ or $A=D/10$, respectively (c.f., Fig.~4 (b),(c)).}
\label{fig:f3}
\end{figure}

The twist instability depends strongly on the geometry of the cell.
In Fig.~2 we plot the numerical results for the 
twist order parameter $N=D^{-3}\int_V{\bf n}_y^2d^3{\bf r}$
as a function of the anchoring strength $W_1$ for cells
with different groove depth $A$.

The free energy of the sinusoidal configuration $F_{sin}$
is estimated by using the variational ansatz
\begin{equation}
\label{eq:alpha}
\alpha_{sin}(x,z) = {\cal B} Aq \cos qx 
\frac{\sinh\lambda z}{\sinh\lambda D} \;.
\end{equation}
This function respects the boundary conditions and
has two parameters ${\cal B}$ and $\lambda$.
If the nematic director field is nearly uniform the dimensionless parameter
${\cal B}\approx 0$. On the contrary, if the director
is parallel to the sinusoidal surface, ${\cal B}=1$.
$\lambda^{-1}$ defines the distance over which the
nematic deformation is influenced by the grating surface.
For small groove depth $A/D$ the total energy is
\begin{equation}
\label{eq:variational_energy}
F_{sin}\approx \frac{S}{4}(Aq)^2
\big[Y(\lambda){\cal B}^2+W_1(1-{\cal B})^2\big]
\end{equation}
where $Y(\lambda)=K_{11}\lambda^2F_1(\lambda)+K_{33}q^2F_2(\lambda)$,
with $F_1=\int_0^D(\cosh\lambda z/\sinh\lambda D)^2dz$ and 
$F_2=\int_0^D(\sinh\lambda z/\sinh\lambda D)^2dz$.
Minimization with respect to the variational parameters yields
$\lambda=q\sqrt{K_{33}/K_{11}}$
and ${\cal B }=W_1/(W_1+Y)$. The total free energy follows as
\begin{equation}
\label{eq:energy0}
F_{sin}\approx \frac{S}{4}(Aq)^2\frac{YW_1}{Y+W_1}\;.
\end{equation}
For $W_1/Y\gg 1$ and $\lambda D\gg 1$ we recover the Berreman energy
\begin{equation}
\label{eq:energy_berreman}
F_{sin}\approx \frac{S}{4}(Aq)^2q\sqrt{K_{33}K_{11}}
\end{equation}
which is independent of the cell size $D$.
However, when the flat surface approaches the grating one
finite size effects come into play.
In this case, for $\lambda D\ll 1$,
$Y\approx[K_{11}+K_{33}(qD)^2/3]/D$.

In order to estimate the critical anchoring strength $W_{1c}$, i.e.,
the threshold for the twist instability, we take
$\phi(x,z)=\phi z/D$ and use
an expansion for the total energy in powers of $\phi$:
\begin{equation}
\label{eq:energy_expansion}
F_t[\alpha,\phi]= F^{(0)}_t[\alpha]+\phi^2F^{(2)}_t[\alpha]+...\;.
\end{equation}
The sinusoidal deformation $\alpha_{sin}$ (with $\phi=0$) minimizes the
free energy if the coefficient $F^{(2)}_t[\alpha_{sin}]$ is positive. 
Supposing that $K_{22}/K_{11}\sim(A/D)^2$ is small one has
\begin{equation}
\label{eq:energy_coefficient}
F^{(2)}_t\approx \frac{SK_{22}}{2D}-\frac{S}{4}(Aq)^2\frac{YW_1}{Y+W_1}\;.
\end{equation}
The twist instability occurs for the threshold value given by 
$F^{(2)}_t[\alpha_{sin}]=0$,
or equivalently when the energy of the sinusoidal deformation equals
\begin{equation}
F^{(c)}_{sin}=\frac{SK_{22}}{2D}\;.
\label{eq:twist_threshold}
\end{equation}
In Fig.~3 we plot the results obtained from solving this equation
for different cell geometries $(A,q=2\pi/L)$.
To favor twist configurations we used $K_{11}=K_{33}=2K_{22}$.

In order to overcome the limitations of the above analytic
analysis extensive numerical calculations of cells with a wide range of
surface parameters (groove depth, pitch, and anchoring strength) have
been carried out. The calculation of the twist transition requires
a numerical procedure capable of evaluating the bulk and 
surface contributions to
the free energy very accurately. Owing to the geometrical pattern of the
grating surface this turns out to be a challenging numerical problem.
We used finite-element discretizations of the functions 
$\alpha(x,z)$ and $\phi(x,z)$ (dividing the space into small triangles
where the functions are approximated linearly) and found the minimum of
the free energy by standard numerical techniques.
Non-uniform adaptive meshes have been used: 
finer meshes were required close to the
grating surface where the fields vary more rapidly \cite{Patricio}. 
In addition, a finite-element triangulation 
that approximates as closely as possible the
geometrical boundary of the cell was required since the
critical anchoring strength depends sensitively on these surface terms.

\begin{figure}[htb]
\par\columnwidth=20.5pc
\hsize\columnwidth\global\linewidth\columnwidth   
\displaywidth\columnwidth
\centerline{\epsfxsize=250pt\epsfbox{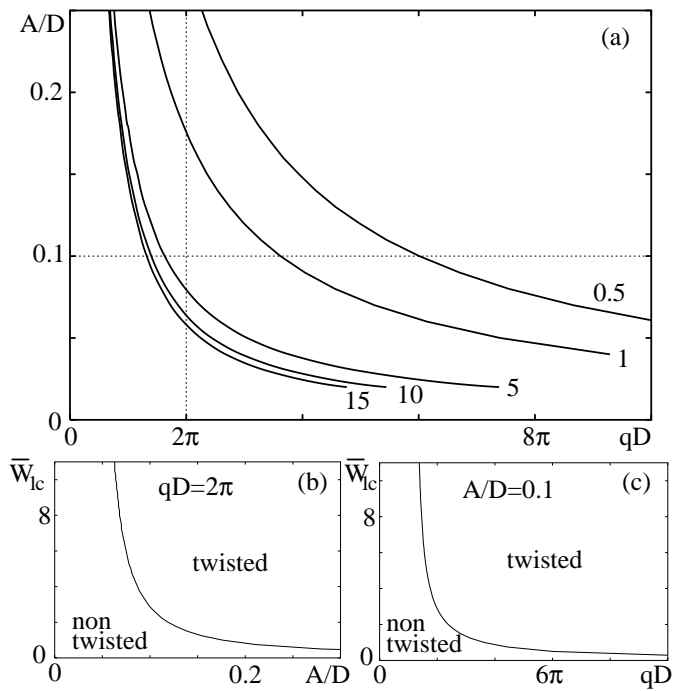}}
\caption
{Critical anchoring strength $W_{1c}(A,q)$ 
for $K_{11}=K_{33}=2K_{22}$ as in Fig.~3
obtained from an exact numerical solution.
(a) Contour lines $\overline W_{1c}=0.5,1,5,10$, and $15$.
(b) $\overline W_{1c}(A,L=D)$. (c) $\overline W_{1c}(A=0.1D,q)$.}
\label{fig:f4}
\end{figure}

These numerical results are shown in Fig.~4.
From this figure one infers that $W_{1c}$ diverges for a set of
the geometrical parameters of the cell. 
This set defines a range of parameters 
for which the nematic orientation will never twist.
Outside this region,
for larger values of $A$ and $q$,
a twisted nematic director may occur for small
values of $\overline W_{1c}=W_{1c}D/K_{11}$.

For completeness, we consider the elastic term associated with
the saddle-splay distortion neglected in the previous analysis. 
This term is usually written in the form
\begin{equation}
\label{eq:energy_k24}
F_{ss}=K_{24} \int_V \nabla \cdot \big[
({\bf n}\cdot\nabla){\bf n}-{\bf n}(\nabla\cdot{\bf n})\big]d^3{\bf r} 
\end{equation}
which reveals its surface nature.
The value of the surface elastic constant $K_{24}$ is bounded for stability
reasons. In fact, the bulk free energy in Eq.(\ref{eq:bulk}) is the sum of
quadratic terms and it is well defined only if the bulk elastic constants
$K_{11}$, $K_{22}$, and $K_{33}$ are positive.
A stability ana\-ly\-sis of the surface energy 
in Eq.(\ref{eq:energy_k24}) yields the constraint
$0<K_{24}<\min(K_{11},K_{22})$ \cite{Schmidt}.
In Fig.~5 we plot the numerical solutions for $\alpha(x,z)$
for physical and unphysical values of $K_{24}$. 
By varying $K_{24}$ (between 0 and $K_{22}$) we found that the 
thresholds $W_{1c}$ for the twist instability are not significantly affected
by this term \cite{Miko} within the physically relevant range.

\begin{figure}[htb]
\par\columnwidth=20.5pc
\hsize\columnwidth\global\linewidth\columnwidth   
\displaywidth\columnwidth
\centerline{\epsfxsize=250pt\epsfbox{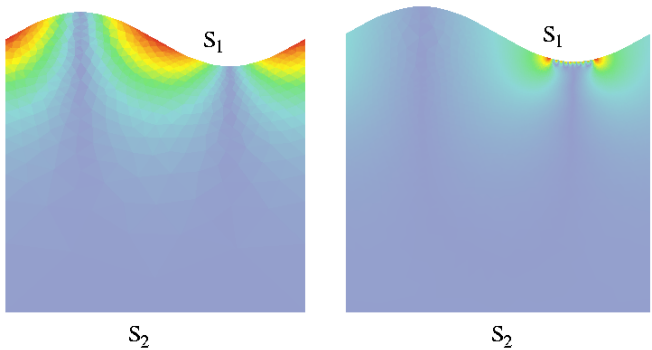}}
\caption
{Numerical solutions for the system considered above 
(with $A/D=0.1$, $qD=2\pi$, $\overline W_1=2$, 
and $K_{11}=K_{33}=2K_{22}$).
The function $|\alpha(x,z)|$ is
represented by a color code in which blue and red correspond
to $\alpha=0$ and $\alpha=\max|\alpha|$, respectively, 
for physical (left: $K_{24}=0$) 
and unphysical values of $K_{24}$ (right: $K_{24}=2K_{11}$).
The unphysical solution on the right exhibits strong, spurious deformations
close to the boundary leading
to a divergent negative free energy that is bounded 
numerically due to the mesh discretization.
The left panel corresponds to the bent sinusoidal configuration
shown in Fig. 1(c).}
\label{fig:f5}
\end{figure}

Beyond the obvious technological importance of the twist cell 
-- it has been shown to possess an electro-optic
response far less dependent on viewing angle than other LC display
configurations \cite{Bryan-Brown} -- the system turns out to be
very interesting also from a theoretical point of view:
for a given anchoring strength, the twist transition is driven in
the absence of an external field by the surface morphology
$(A/D,qD)$. Close to the surface transition, an arbitrarily small 
external field or temperature variation
will be sufficient to induce the reorientation of the nematic.

An obvious extension of this work includes application of a voltage
between conducting substrates. This problem, however, requires special
care.
At constant voltage the system is no longer isolated
and the {\sl minimum} principle for the total free energy 
(with electric and elastic terms) does not apply. The free energy is
minimal with respect to the nematic director field and maximal  
with respect to the elecric potential \cite{Landau}. 
A generalization of the numerical method described above may not converge
and a numerical solution of the Lagrangian
differential equations appears to be required.

PP acknowledges the support of the Funda\c c\~ao para a 
Ci\^encia e Tecnologia (FCT) through Grant No. SFRH/BPD/5664/2001.

\end{multicols}

\end{document}